\documentclass[prb,preprint]{revtex4-1} 

\usepackage{xcolor}
\usepackage{amsmath}  % needed for \tfrac, \bmatrix, etc.
\usepackage{amsfonts} % needed for bold Greek, Fraktur, and blackboard bold
\usepackage{graphicx} % needed for figures
\usepackage[utf8]{inputenc}
\usepackage{gensymb}
\begin{document}

\title{Using mobile-device sensors to teach students
    error analysis}

\author{Martín Monteiro}
\email{monteiro@ort.edu.uy} % optional
\affiliation{Universidad ORT Uruguay}

\author{Cecila Stari}
\author{Cecila Cabeza}
\author{Arturo C. Martí}
\email{marti@fisica.edu.uy}

\affiliation{Instituto de F\'{i}sica, Facultad de Ciencias,
  Universidad de la Rep\'{u}blica, Igu\'{a} 4225, Montevideo, 11200,
  Uruguay}
\date{\today}

\begin{abstract}
  Science students must deal with the errors inherent to all physical
  measurements and be conscious of the need to expressvthem
  as a best estimate and a range of
  uncertainty. Errors are routinely classified as statistical or
  systematic. Although statistical errors are usually dealt with in
  the first years of science studies, the typical approaches are based
  on manually performing  repetitive observations.
  Our work proposes a set of laboratory experiments to teach
  error and uncertainties based on data recorded with the sensors available in many mobile devices. The main aspects addressed are
  the physical meaning of the mean value and standard deviation, and
  the interpretation of histograms and  distributions. The normality of the fluctuations is analyzed qualitatively comparing  histograms with normal curves and quantitatively comparing the number of observations in  intervals to the number expected  according to a normal distribution
  and also performing a Chi-squared test. We show that the distribution usually follows
  a normal distribution, however, when the sensor is placed on top of a loudspeaker
  playing a pure tone significant differences with a normal distribution are observed. 
As applications  to every day situations we discuss the intensity of the fluctuations in different situations, such as placing the device on a table or holding it with the hands in different ways. Other activities are focused on the smoothness of a road quantified in
terms of the fluctuations registered by the  accelerometer. The present proposal contributes to  gaining a deep insight into modern technologies and statistical
errors and, finally, motivating and encouraging engineering and
science students.
\end{abstract}

\maketitle

\section{Introduction}
In many experimental situations when a measurement is repeated --for
example when we measure a time interval with a stopwatch,  the
landing distance of a  projectile or a voltage with a digital multimeter--
successive readings give slightly different results  under identical conditions.
This occurs beyond the care we take to always launch the
ball exactly the same way or to connect the components of the
circuit so that they are firmly attached. In effect, this phenomenon
occurs in the real world because most measurements present statistical uncertainties
%\cite{taylor1997introduction,bevington1993data,hughes2010measurements}
\cite{taylor1997introduction,hughes2010measurements}.  When facing
repeated observations with different results it is natural to ask
ourselves which value is the most representative  and what confidence
level   can we have in that value.  The International Standard
Organization (ISO) \cite{iso1995guide} (see also
  \cite{taylor2007nist,bureau2008evaluation}) defines the errors
evaluated by means of the \textit{statistical} analysis of a series of
observations as type A,  in contrast with other sources of errors that are \textit{systematic} and defined as type B. The  evaluation of the latter
is estimated using all available non-statistical information such as instrument
characteristics or the individual judgment of the observer. In this work, we
focus on the teaching of statistical errors in the first years of
science and engineering studies using modern sensors.

The study of error analysis and uncertainties plays a prominent role
in the first years of all science courses.On this
  matter, AAPT recommends \cite{kozminski2014aapt} that students
  \textit{should be able to use statistical methods to analyze data
    and should be able to critically interpret the validity and
    limitations of the data displayed. In general, a physicist must
  be able to design a measurement procedure, select the equipment or
  instruments, perform the process and finally express the results as
  the best estimate and its of uncertainty}.  Perhaps the most important
message is to persuade students that any measurement is useless unless
a confidence interval is specified.  It is expected that, after
finishing their studies, students are able to discuss whether a result
agrees with a given theory and, if it is reproducible, or to distinguish a
new phenomenon from a previously known one. With this objective, various experiments are usually proposed in introductory laboratory courses
\cite{4321968,mathieson1970student,fernando1976experiment,arvind2004random,wibig2013hands,gan2013simple}. These
experiments usually involve a great amount of repetitive measurements
such as dropping small balls \cite{gan2013simple}, measuring the
length of hundreds or thousands of nails using a
vernier caliper \cite{fernando1976experiment} or randomly sampling an alternating current source  \cite{arvind2004random}.  The measurements obtained are usually
examined from a statistical viewpoint plotting, histograms,
calculating mean values and standard deviations and, eventually,
comparing them with those expected from a known distribution, typically a
normal distribution.  Although these experiments are illustrative, most
of them are tedious and do not adequately reflect the present state of
the art.

In contrat with the expectable learning outcomes mentioned above,
recent studies  \cite{Sere1993learning,allie2003teaching,chimeno2005teaching} report
  several difficulties associated with error analysis among students.  The
  study by S\'er\'e \textit{et al} \cite{Sere1993learning} highlighted
  the lack of understanding of the need to make several measurements,
  the poor insight into the notion of confidence intervals or the inability to distinguish between random and systematic errors. Another
  investigation \cite{allie2003teaching} remarked the inconsistency of the
  common  view of students with generally accepted scientific models. On
  many occasions, the student's model of  thinking is close to a
  \textit{point} paradigm as opposed to a more elaborated
  \textit{probabilistic} interpretation of the measurements.  A
  reseach-based assesment showed that although the impact of
  introductory laboratory courses was positive, only a relatively small
  percentage of students developed a deeper understanding of
  measurement uncertainty \cite{PhysRevSTPER.4.010108}. 

The fluctuations present in the sensors of modern
mobile devices give rise to an alternative approach to teaching error
analysis. Indeed, most smartphones and tablets come equipped with
several built-in sensors such as accelerometers, magnetometers,
proximeters or ambient-light sensors. Several Physics experiments using these sensors have been proposed in recent years (see for
example \cite{vieyrafive,hochberg2018using,monteiro2016using}).  In
  almost all the experiments, only mean values are taken into
  consideration; however, due to their sensitivity, sensor readings
  also display statistical fluctuations. Although being detrimental in many
  situations, these fluctuations can be used favorably to illustrate
  basic concepts related to the statistical treatment of measurements. 
  Using these sensors,  it is certainly possible to acquire hundreds or
  thousands of repeated values of a physical magnitude in a few
  seconds and analyze them in the mobile device or in a PC. We propose
  here a set of laboratory activities to teach error analysis and
  uncertainties in introductory Physics laboratories based on the
  fluctuations registered by mobile-device sensors. In the next
  Section we describe the basic set of activities while Section
  \ref{sec:other} is focused on other applications that take into
  account sensor fluctuations in non-standard situations. Finally, in
  Section~\ref{sec:con} we present the summary and conclusion.

\section{A laboratory based on mobile devices}
\label{sec:lab}
Among all the sensors, accelerometers, capable of
  measuring the acceleration of the device in the three independent
  spatial directions, are the most ubiquitous in mobile
  devices. Though it is possible to use accelerometers or anyone of
  the others or even more than one sensor simultaneously here, for
  the sake of clarity, the proposed experiments are mainly based on
  the $z$ component of the acceleration $a_z$, defined as
  perpendicular to the screen.  As a general rule, the characteristics
  of the sensors can be found using specific applications (\textit{apps}) or looking for datasheets
  in the internet. The range (difference between the maximum and minimum
  value that it is capable of measuring) and the resolution  (minimum
  difference that the sensor can register, which is sometimes  incorrectly termed 
  as accuracy) of several sensors are summarized in
  Table~\ref{tabsensor}.  It is worth remarking that, although being
  universally known as accelerometers, in fact, they are
  \textit{force} sensors \cite{monteiro2014exploring,MONTEIRO2015}.
  Indeed, a device standing on a table would register a value close to
  the gravitational acceleration in the vertical axis while a free-falling device would register a value close to zero in the same axis.  

\begin {table}[h]
\begin{center}
\begin{tabular}{|c|c|c|c|}  
\hline 
Phone &	Sensor	&  Range (m/s$^2$)	&  Resolution (m/s$^2$)    \\  \hline 
Samsung Galaxy S7	 &K6DS3TR&	$\pm78.4532$	&0.0023942017   \\ 
LG G3	&LGE&	$\pm39.226593$	&0.0011901855   \\ 
Nexus 5	&MPU-6515	&$\pm19.613297$&	0.0005950928   \\ 
 iPhone 6&	MPU-6700	& - &	-    \\ 
Samsung J6+&	LSM6DSL&	$\pm39.2266$ &	0.0011971008   \\ 
Xiaomi Redmi Note7	&ICM20607	&$\pm78.4532$&	0.0011901855   \\ 
Samsung Galaxy S9	&LSM6DSL	&$\pm78.4532$&	0.0023942017   \\ 
Samsung A20S	&ICM40607&	$\pm78.4532$	& 0.0023956299 \\
\hline
\end{tabular}  
\caption{\label{tabsensor} Range and resolution of
    several common mobile devices obtained with the Androsensor
    app. Notice that sensors  can usually operate within different ranges
    $\pm 2g$, $\pm 4g$, $\pm 8g$,..., which are set by the
    mobile-device manufacturer. The resolution depends on the choice
    of range.}  In the case of the iPhone the manufacturer does
  not provides this information.
\end{center}\end {table}

\begin{figure}[ht]
\begin{centering}
\includegraphics[width=0.895\columnwidth]{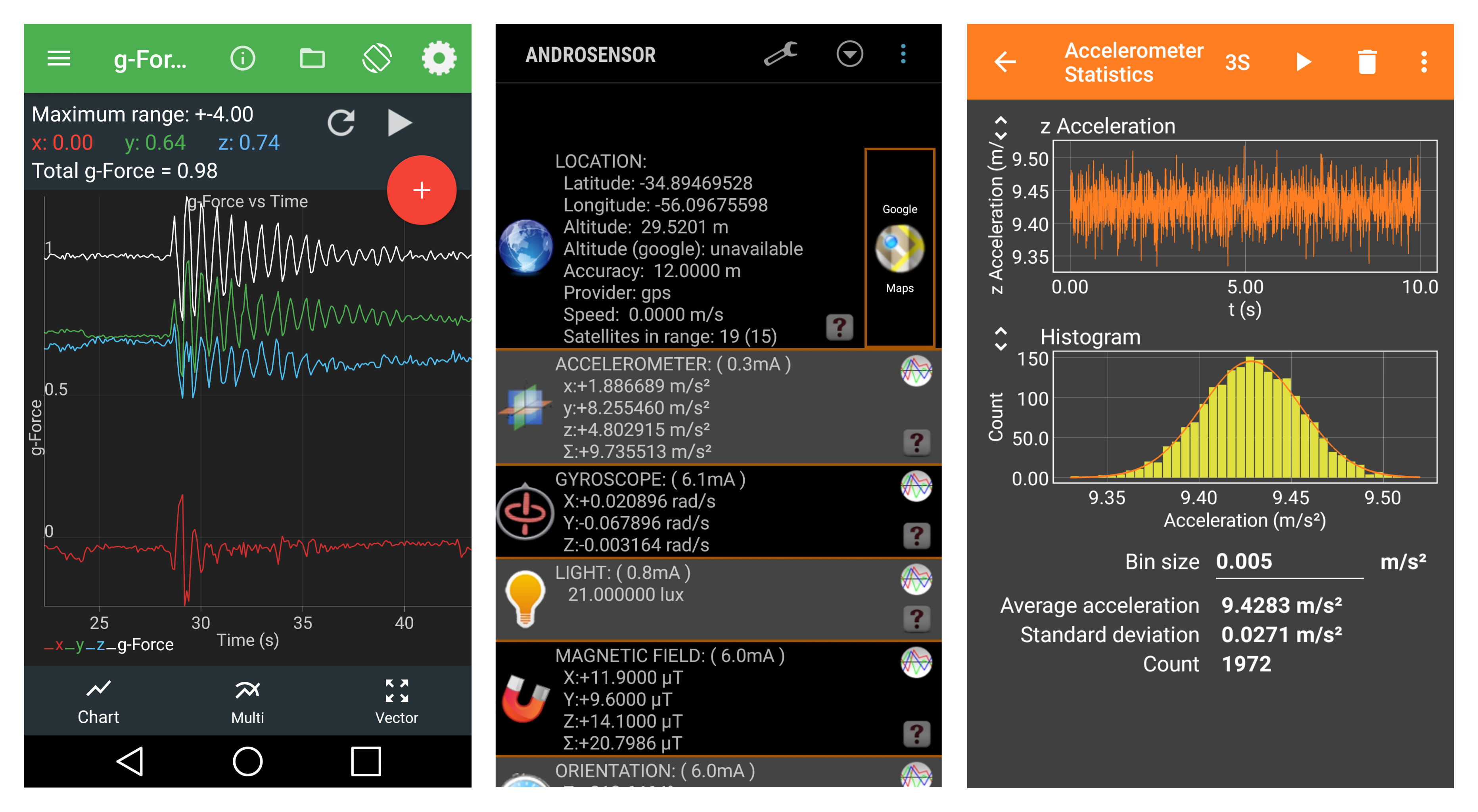}
\caption{\label{fig2apps} Screenshots of three suitable \textit{apps}: Physics Toolbox suite (left), Androsensor (center), Phyphox (right). The right panel shows a Phyphox screenshot of the experiment \textit{Statistical Basics} including a temporal series of the vertical component of the acceleration (top) and the corresponding
histogram (bottom) overlapped with a Gaussian curve with the same mean and standard deviation indicated in the image.}
\end{centering}
\end{figure}

In general, a specific piece of software, or an app, is
necessary. Digital stores offer many apps that are able to communicate with the sensors.
In particular, Physics Toolbox Suite \cite{physicstoolbox}, Androsensor
and PhyPhox \cite{Staacks_2018}, whose screenshots are shown in
  Fig.~\ref{fig2apps}, are suitable for the experiments proposed
  here. Using these \textit{apps} it is possible to select the
relevant sensors, and to setup the parameters such as the duration of
the time series and the sampling frequency.  The registered data can
be analyzed directly on the smartphone screen or transferred to the
cloud and studied on a PC using a standard graphics package.  
Others useful characteristics of these apps are the delayed execution and the remote access via \textit{wi-fi} or browser. These capabilities allow the experimenter to avoid  touching or pushing
the mobile device once the experiments has started.

\subsection{Normal distribution of the sensors'  fluctuations}

The first experiment consists of recording the fluctuations of the
vertical component of the accelerometer sensor with the mobile device
in three different situations: laid on a table, hand-held 
and resting on another smartphone playing a 600 Hz pure
tone. In all the cases, we choose, unless stated otherwise, a delay of
$3$ s and register $a_z$ for $30$ s.  The delay is important in order to avoid
touching the device when the register starts and thus introducing
spurious values.  Let us denote $N$ the number of
  measurements registered by the sensor, $\overline{a}_z$ the mean
  value and $ \sigma_{a_z}$ the standard deviation.  

\begin{figure}[h]
\begin{centering}
\includegraphics[width=0.99\columnwidth]{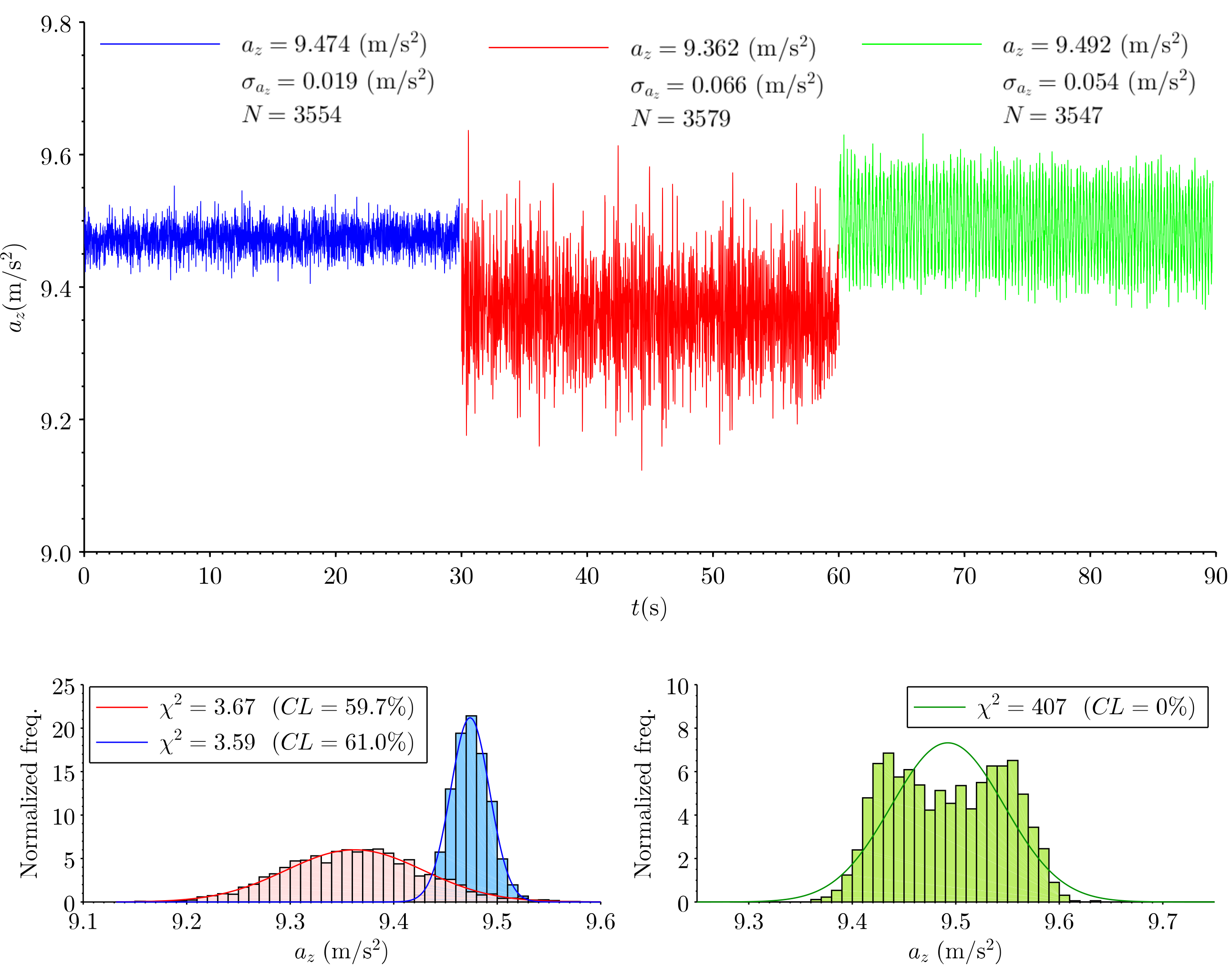}
\par\end{centering}
\caption{\label{figA} Fluctuations registered by the accelerometer. The top
  panels display $a_z(t)$ with the device horizontal in three
  different situations: laid on a table (blue), hand-held 
  (red) and resting on another smartphone playing a 600 Hz pure tone
  (green). The smartphone was the LG-G3 with a  $\Delta t=0.004$ s sampling period.
  The bottom panels display the histograms and the continuous lines with the same color are normal (Gaussian) functions   with same mean value, standard deviation and normalization. Legend
  boxes indicate $\chi^2$ values, calculated with 8 bins, and the
  corresponding confidence levels (CL).}

\end{figure}

The results of the experiment are summarized in
  Fig.~\ref{figA} in which the top panels display the temporal series,
  $a_z(t)$, and the bottom panels show the histograms using 
  the same respective colors. In all the cases the accelerations
  fluctuate stationarily around  mean values.  Although these
  values are close to the well-known value of the gravitational
  acceleration, they are slightly different and they are not expected
 to represent a measure of that magnitude. This is due to several 
 reasons, for instance it depends on the horizontalitly of the
  table or the hand or also on the calibration of the sensor. It is
  interesting for students to check that the mean value changes when
  the device is laid on a table with the screen pointing upwards
  or downwards. Another possible, and equivalent, alternative (not
  shown here) consists of plotting $a_x(t)$ or $a_y(t)$ which exhibit
  similar temporal evolutions and histograms but fluctuating around a
  value close to $0$ m/s$^2$.

The differences in the intensity of the fluctuations
  exhibited in the three mentioned situations are evident in the
  top panels of Fig.~\ref{figA}.  The intensity is clearly larger when
  the smartphone is hand-held (red) or under the influence of
  the 600Hz tone (green) in comparison with the smartphone on a table
  (blue). The standard deviation of each series, indicated in the
  legend boxes, is clearly related to the intensity of the
  fluctuations.  This observation substantiates the use of the
  standard deviation in the framework of the applications proposed in
  Section \ref{sec:other}.

A relevant aspect to study is the distribution of
the fluctuations and how it compares with the normal distribution. 
The firt approach to testing the normality of the distribution
is qualitative. In the bottom   panels (Fig.~\ref{figA}), the histograms are compared with normal (Gaussian) functions with the same mean values and standard deviation
  and the vertical scale adjusted so that the area under the normal
  curve and the sum of the bins of the histogram are equal.
It can be observed that the histograms and normal functions
  agree very well in the cases of the blue and red curves. By
  increasing the number of samples $N$ and simultaneously decreasing
  the width of the bins, it is possible to see that the
  agreement improves even more  (not shown here).  Contrarily, in the green case, the
  pure tone \textit{breaks} the normality of the distributions as it
  is clearly revealed by the disagreement between the histogram and
  normal curve.

The second and more quantitative approach to verifying the normality of the distributions is given by the comparison of the fraction of
observations in a given interval around the mean value and the
expected percentage according to a normal distribution.
Table~\ref{tab:sigma} displays these percentages for the 
experiment depicted in Fig.~\ref{figA}.  It can be seen that, in
agreement with the qualitative test, the observed and expected percentages
are quite similar when the device is on the table or hand-held  where they present considerable divergences under the influence of the 600 Hz tone.
\begin {table}[ht]
\begin{center}
\begin{tabular}{|c|c|c|c|c|} \hline 
Experiment   &	Table &	Hand	 &Speaker &Theoretical \\  \hline
N & 3554	& 3579	& 3547 &- \\ \hline
$\overline{x}\pm \sigma$ (m/s$^2$) &  $9.474 \pm0.019$	&$9.362 \pm 0.066$ &$9.492 \pm0.054$ & - \\ \hline
$(\overline{x} - \sigma, \overline{x}+ \sigma)$ &	69.3\% &67.9\%	&58.2\%  &	68.2\%\\
$(\overline{x} - 2\sigma, \overline{x}+ 2\sigma)$   &	95.2\% &95.5\%	&99.2\%  &	95.4\%\\
$(\overline{x} - 3\sigma, \overline{x}+ 3\sigma)$   &	99.7\% &99.7\%  &100.0\%  &	99.7\%\\ \hline
\end{tabular}
\caption{\label{tab:sigma} Fraction of observations in intervals
  around the mean defined in units of the standard deviation compared
  with the expected number according to a normal distribution. Each
  column corresponds to each of the temporal series plotted in
  Fig.~\ref{figA}.}
\end{center}
\end {table}

To gain further insight into the normality of the distributions, a
chi-squared test comparing the difference between the number of observations
measured and expected in each bin \cite{taylor1997introduction}, was
performed. Each $\chi^2$ valued can be associated with a confidence
level that determines the rejection of the hypothesis of normal
distribution. Clearly in the  blue and red cases the $\chi^2$ test
indicates the compatibility of the normal distribution hypothesis 
while in the green case this hypothesis must be rejected.

\subsection{Resolution in digital sensors}
\begin{figure}
\begin{centering}
\includegraphics[width=0.8\columnwidth]{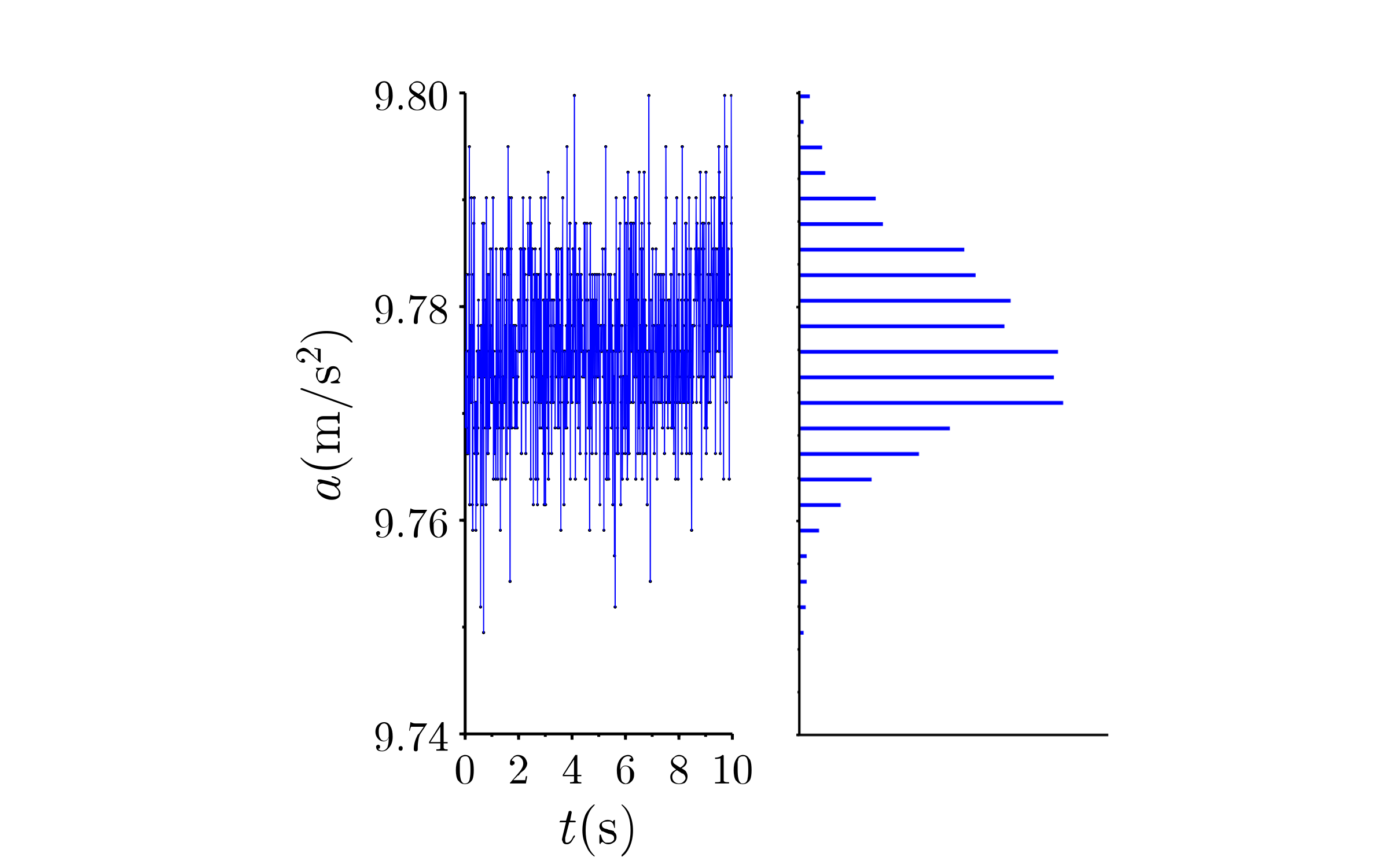}
\par\end{centering}
\caption{\label{figC} Discrete nature of the sensor data. The horizontal
  axis in the left panel was zoomed out to emphasize the discrete
  nature of the accelerometer values. The right panel shows the same
  values in a horizontal histogram with the same
  vertical scale.}
\end{figure}

By zooming in on the temporal series displayed in Fig.~\ref{figA}, 
it can be seen that the sensor values do not take continuous values, but only a discrete set is possible. This is more evident in the experiment displayed in Fig.~\ref{figC} where the horizontal axis 
has been zoomed out  in the left panel, and a horizontal histogram with the same values 
is shown in the right panel. The difference between the discrete values in the vertical axis is the
resolution of the instrument, that is, the minimum difference that the
sensor can register. This is typical of digital instruments, where a
continuous magnitude (such as acceleration, in this case) is
transformed by a sensor into an analog electrical signal, which is in turn
transformed by an analog-to-digital converter (ADC) into a digital
signal which can only take certain discrete values. In this case, the
acceleration sensor of the Samsung S7 is a K6DS3TR and its resolution,
indicated in Table~\ref{tabsensor}, is $\delta = 0.0023942017$ m/s$^2$
which corresponds exactly to the difference between consecutive
acceleration values.

The resolution of the sensor, $\delta$, is the
  quotient between the range, $2R$, and the number of different values
  that the sensor can register, $2^n$,
\begin{equation}
 \delta =\frac{2 R}{2^n}
\end{equation}
where $n$ is the number of bits of the sensor and the factor $2$
stands because it registers not only positive measures,
but also negative accelerations. Taking into consideration Table
\ref{tabsensor}, it can be determined that this sensor is capable of
measuring $2 R/ \delta = 65536$ different values and since
$65536 = 2^{16}$, this means that it is a $16$-bit sensor, which can be
easily verified on the data sheets.

\subsection{Standard error and optimal number of measurements}

The standard deviation, if $N$ is large enough, is characteristic of
the set of all the possible observations whereas the standard error, or standard deviation of the mean, generally defined as $\sigma(\bar{a_z}) = \sigma_{a_z}/\sqrt{N}$  represents the margin of uncertainty of the mean value obtained in a particular
set of measurements \cite{taylor1997introduction}. The result of a
specific measurement is usually expressed as
$\overline{a}_z \pm \sigma(\bar{a_z})$
representing the best estimate and the confidence in that value.  In
Table.~\ref{tabsigmaN} the standard deviation is shown as a function of $N$.
As mentioned above, it is clear from that data that $\sigma_{a_z}$ is nearly
constant and, as a consequence, $\sigma(\bar{a_z})$ is proportional to
$N^{-1/2}$.

\begin{table}
 \begin{center}
\begin{tabular}{|c|c|c|c|c|c|c|c|c|c|c|} \hline 
$N$ &563 & 1156  &  1746 &  2348 & 2941 &3535 & 4166 &4733& 5327& 5919\\ \hline
$\sigma_{a_z}$(m/s$^2$) &0.020&0.019 & 0.018 & 0.019 & 0.020& 0.019 & 0.019 & 0.019 & 0.019 &0.020\\ \hline
\end{tabular}
\caption{\label{tabsigmaN}Standard deviations of $a_z(t)$ corresponding to several experiments under identical conditions but with different number of measurements.}
\end{center}
\end{table}
The choice of $N$ in a specific experiment is a delicate question.  Indeed, if we could repeat the measurements infinite times the standard error would vanish and we could achieve a perfect knowledge of the best estimate.
However, as the decrease of the standard error with the number
of observations is slow, it is impractical to increment 
this number excesively. A common criterion  is to take a number of measurements,
often referred as the \textit{optimal number of measurements}, $N_{\mathrm{opt}}$,
such that the statistical uncertainty is of the same order as the 
systematic  (or type B) errors. Here, in the absence of other sources of
systematic errors, the standard error should be of the same order as the resolution of the digital instrument: $\sigma(\bar{a_z}) = \sigma_{a_z} /\sqrt{N_{opt}} \sim \delta$. In the  experiment depicted in Table~\ref{tabsigmaN} with a LG G3, the resolution is  $\delta=0.0012$ m/s$^2$, therefore $N_{\mathrm{opt}} \sim 250.$

\section{Other applications}
\label{sec:other}
In this Section we propose a couple of activities in which the knowledge of the fluctuations measured by a sensor can contribute to quantify another magnitude.

\subsection{The steady hand game}

An interesting experiment is to study the intensities of the
fluctuations depending on the way in which the experimenter holds
his/her device.  This activity can be adapted for a group of students  as a
challenge  consisting of trying to hold the
device as steadily as possible.  Another possibility (not recommended
by the authors) is to study the fluctuations of the gait of a pedestrian as a function of the
alcohol beverage intake similar to \cite{zaki2020study}.

\begin{figure}
\includegraphics[width=0.8\columnwidth]{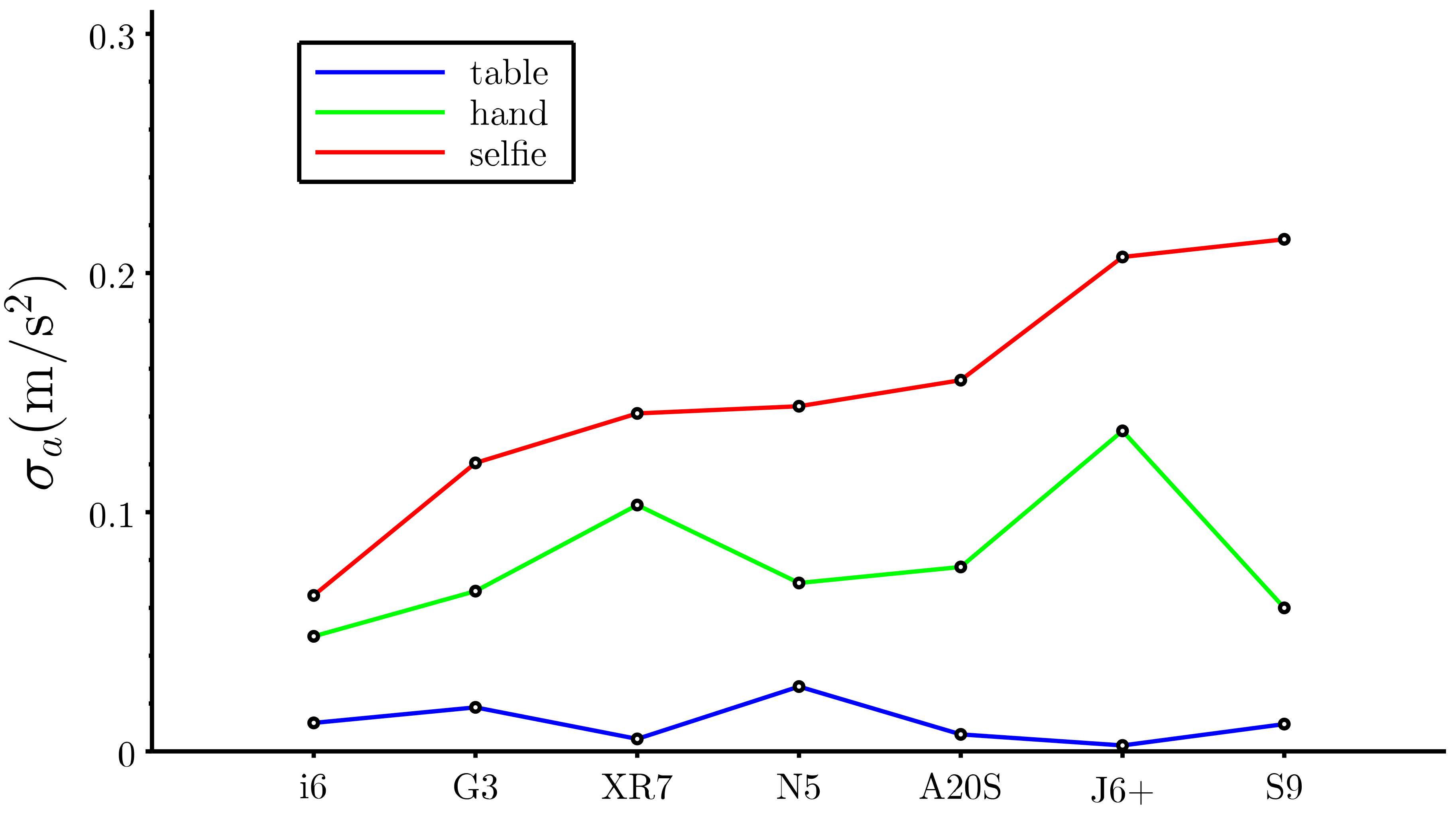}
\caption{\label{figsigma} Comparative table of the standard
  deviation $\sigma$ for different mobile devices in different
  activities as a function of the different models (see
  Table~\ref{tabsensor}). Lines are guides for the eyes}
\end{figure}

The steadiness of the device is quantified by the standard deviation
of a given temporal series. The intensities of the fluctuations in different situations and for different sensors are displayed in Fig.~\ref{figsigma}.
It is evident from these values that the mobile device on the table exhibits in all the cases less fluctuations than when the device is held by the experimenter.  Moreover 
a more stable position is achieved by keeping the device close to the trunk as opposed to the classical \textit{selfie} position. 
Another point worth mentioning  is that the intensity of the fluctuations depends on the specific sensor but exhibits in all cases the same trends mentioned above. 
Several interesting extensions to this experiment can be proposed:
the dependence on the characteristics of the experimenter (age, training, concentration). The origin of the fluctuations, mechanical or electronical,
can be considered in case of having  an antivibration table. When using interferometric methods a precise calibration of the mobile device sensor could  also be performed.

\subsection{The smartphone as a way to assess  road quality}
Recently, smartphones' sensors were proposed to assess road quality
\cite{harikrishnan2017vehicle}. In this activity, which can be
performed outdoors, students can assess the quality of a road. A means
of transport, in this case a car,  is employed under similar conditions (speed), but other possibilities, such as a bike, are equally
feasible.  The intensities of the fluctuations traveling by car on different roads are listed
in Table~\ref{tab:roads}. To get an insight of the fluctuations attributable  to the road
the noise with the car stopped and the engine idle is indicated in the first road.
A similar measurement  performed in a flying aircraft is
included solely for the sake of comparison.
This activity can be extended to evaluate comfort in any other
means of  transport media, for example, elevators.

\begin {table}[ht]
\begin{center}
\begin{tabular}{|l|l|l|l|l|} \hline 
 Situation & N & $\sigma_\mathrm{G3}$ (m/s$^2$) & N& $\sigma_\mathrm{XR7}$ (m/s$^2$)\\ \hline
Engine idle     &1181 & 0.3818 & 4984 & 0.0352\\
Smooth pavement &1200 & 1.3487 & 4974 & 0.5642\\
Stone pavement  &   -  &  -    & 4952 & 1.1491\\ 
Aircraft        & 1999 & 0.4374& -    & - \\ 
\hline
\end{tabular}  
\caption{\label{tab:roads} Assessment of the quality of different
  roads. Standard deviation of $a_z$ while the device is on the floor
  of the car with the screen orientated upwards.}
\end{center}
\end {table}

\section{Summary and conclusion}
\label{sec:con}
The activities discussed above were successfully proposed at our university to freshman Engineering and Physics students. In previous laboratory experiments students already knew the usefulness of the sensor to study several phenomena
in which the noise was a factor to avoid. However, 
when sensors were proposed to study fluctuations, they were surprised
to discover a kind of \textit{underlying world}. Despite having gone
through statistical topics in several courses, the normal distribution
appearing as an experimental fact rather than a mathematical consequence, was original. It is worth discussing, how the distributions of the other sensors
change. For example, magnetometer fluctuations, significantly influenced by  motors or ferromagnetic material in the vicinity, do not follow normal distributions.  Several challenges can be proposed
in relation to sports, comfort evaluations  or quality control.

The main conclusion is that modern mobile-device sensors are useful
tools for teaching error analysis and uncertainties. In this work we
proposed several activities that can be performed to teach
uncertainties and error analysis using digital instruments and the
builtin sensors included in modern mobile devices. It is straightforward to obtain experimental distributions of fluctuations and compare them with the expected ones. 
It is shown that the distributions  usually obey normal (Gaussian)
statistics; however, it is easy to obtain non normal distributions. 
The role of noise intensity, spreading or narrowing the distributions open up the possibilities of new applications. Holding the mobile device in different ways also gives an idea of how firmly it is held. Registering acceleration values in a car can assess the smoothness of a road. In this approach, the lengthy and laborious manipulations of
traditional approaches based on repetitive measurements, are avoided allowing teaching to focus on the fundamental concepts. These experiments could contribute to motivating students and showing them the necessity of considering uncertainty analysis.
Several possible extensions related to non-normal statistics can be considered,
such as Poison distribution \cite{mathieson1970student}, distribution of maxima, Chauvenet criterion \cite{limb2017inefficacy}, or Benford law \cite{bradley2009benford}. 

\section*{Acknowledgment}
The authors would like to thank PEDECIBA (MEC, UdelaR, Uruguay)
and express their gratitude for the grant Fisica Nolineal (ID 722)
Programa Grupos I+D CSIC 2018 (UdelaR, Uruguay).

\bibliography{/home/arturo/Dropbox/bibtex/mybib}

\bibliographystyle{unsrt}

\end{document}